\documentclass[%
 reprint,
 amsmath,amssymb,
 aps,jcap,
]{revtex4-2}
\usepackage{natbib}
\usepackage{newtxtext,newtxmath}

\usepackage[T1]{fontenc}
\usepackage{blindtext}
\usepackage{color}
\usepackage{xspace} 

\definecolor{ForestGreen}{rgb}{0.13, 0.55, 0.13}
\definecolor{awesome}{rgb}{1.0, 0.13, 0.32}
\definecolor{amethyst}{rgb}{0.6, 0.4, 0.8}

\newcommand{\kmax}{k_{\rm max}}

\newcommand{\cGpc}{\, h^{-3} \, {\rm Gpc}^3}
\newcommand{\kMpc}{\, h \, {\rm Mpc}^{-1}}
\newcommand{\kreach}{k_{\rm reach}}

\newcommand{\PL}{P_{\rm L}}

\newcommand{\kv}{\mathbf{k}}

\newcommand{\qv}{\mathbf{q}}

\newcommand{\de}{\mathrm{d}}
\newcommand{\sancho}{\textsc{Sancho}\xspace}
\newcommand{\quijote}{\textsc{Quijote}\xspace}
\newcommand{\pbj}{\textsc{PBJ}\xspace}
\usepackage{graphicx}	
\usepackage{amsmath}
\usepackage{hyperref}
\usepackage[mathlines]{lineno}
\usepackage{aas_macros}

\begin{document}

\preprint{APS/123-QED}

\title[On the range of validity of perturbative models]{On the range of validity of perturbative models for galaxy clustering and its uncertainty}

\author{Giosuè Gambardella,$^{1,2}$ \thanks{E-mail:giosak98@gmail.com} Matteo Biagetti,$^{3,1,4}$ 
Chiara Moretti,$^{4,7,1,5}$} \author{Emiliano Sefusatti$^{5,1,6}$} 

\affiliation{$^{1}$Institute for Fundamental Physics of the Universe, Via Beirut 2, 34151 Trieste, Italy}
\affiliation{$^{2}$Institute of Space Sciences (ICE, CSIC), Campus UAB, Carrer de Can Magrans, s/n, 08193 Barcelona, Spain}
\affiliation{$^{3}$Area Science Park, Localit\`a Padriciano 99, 34149 Trieste, Italy}
\affiliation{$^{4}$Scuola Internazionale di Studi Superiori Avanzati, via Bonomea 265, 34136 Trieste, Italy}
\affiliation{$^{5}$Istituto Nazionale di Astrofisica, Osservatorio Astronomico di Trieste, via Tiepolo 11, 34143 Trieste, Italy}
\affiliation{$^{6}$Istituto Nazionale di Fisica Nucleare, Sezione di Trieste, via Valerio 2, 34127 Trieste, Italy}
\affiliation{$^{7}$Centro Nazionale ``High Performance Computer, Big Data and Quantum Computing''}

\date{\today}

\begin{abstract}
We explore the reach of an analytical model at one-loop in Perturbation Theory (PT) to accurately describe measurements of the galaxy power spectrum from numerical simulations in redshift space. We consider the validity range in terms of three different diagnostics: 1) the goodness of fit; 2) a figure-of-bias quantifying the error in recovering the fiducial value of a cosmological parameter; 3) an internal consistency check of the theoretical model quantifying the running of the model parameters with the scale cut.  We consider different sets of measurements corresponding to an increasing cumulative simulation volume in redshift space. For each volume we define a median value and the associated scatter for the largest wavenumber where the model is valid (the $k$-reach of the model). We find, as a rather general result, that the median value of the reach decreases with the simulation volume, as expected since the smaller statistical errors provide a more stringent test for the model. This is true for all the three definitions considered, with the one given in terms of the figure-of-bias providing the most stringent scale cut.  More interestingly, we also find that the error associated with the $k$-reach value is quite large, with a significant probability of being as low as 0.1$\kMpc$ (or, more generally, up to 40\% smaller than the median) for all the simulation volumes considered. We explore as well the additional information on the growth rate parameter encoded in the power spectrum hexadecapole, compared to the analysis of monopole and quadrupole, as a function of simulation volume. While our analysis is, in many ways, rather simplified, we find that the gain in the determination of the growth rate is quite small in absolute value and well within the statistical error on the corresponding figure of merit. 
\end{abstract}
\maketitle
\section{Introduction}

The theoretical modelling of the galaxy power spectrum in redshift space has witnessed, over the past two decades, significant advancements in terms of an accurate description, at the one-loop level, of nonlinear matter evolution, galaxy bias, and redshift-space distortions (see, e.g. \citep{AlamEtal2017, IvanovSimonovicZaldarriaga2020, DAmicoEtal2020, SemenaiteEtal2022} for some recent analyses). The range of scales included in the analysis of a given power spectrum measurement can be limited by the reach of the theoretical model, determined by the perturbative order of the model and, ultimately, by distinct characteristic scales determined by the nonlinear matter evolution, non-local galaxy bias, redshift-space distortions as well as possible corrections to Poisson shot-noise. 

Despite some valuable efforts \citep{BaldaufEtal2016, ChudaykinIvanovSimonovic2021}, the various and complex physical processes at play do not allow a simple but accurate estimate of theoretical error due to neglecting higher-order corrections. For this reason, in practice, the range of validity of the model is determined in actual data analysis by comparing results varying the maximum wavenumber $k_{\rm max}$ making sure the posteriors on cosmological parameters are not running. In \citep{ZhangEtal2022, DAmicoEtal2022B} the effect on parameters constraints of representative next-order corrections is also considered in the determination of $k_{\rm max}$. This of course still depends, inevitably, on the specific realisation of the galaxy and matter distributions. Alternatively, the "scale cut" could be determined by comparison to N-body simulations (see \citep{IvanovEtal2023, DAmicoEtal2020}). The scale-cut will then be applied to a data set again affected by statistical scatter. The question is then how conservative should this choice be to account for such scatter. All these approaches are affected, in different ways, by statistical errors, in addition to other limitations, as, for instance, the reliability of N-body simulations and the synthetic galaxy catalog.  

In this paper, we provide a simple numerical exploration of this issue. Specifically, we study how the range of validity of a one-loop PT model for the galaxy power spectrum multipoles in redshift space, determined from a single measurement over a given volume, is affected by the statistical error. We do so considering a very large set of simulated galaxy catalogs, the \sancho data set, and considering the mean of power spectrum measurements from subsets of increasing size, therefore corresponding to an increasing effective volume. For each value of the subset cumulative volume, we derive scale cut estimates from all available subsets, and therefore an estimate of its scatter. We proceed to study these results as a function of the cumulative volume. 

This paper is organised as follows. In section~\ref{sec:data} we present the simulations data set, the power spectrum measurements, and their covariance. In section~\ref{sec:methodology} we present the methodology, with a brief mention of the theoretical model and parameters implemented in the inference pipeline, and a description of the Likelihood analysis. In section \ref{sec:results} we present our results and conclusions.

\section{Data and measurements}
\label{sec:data}

\paragraph*{The \sancho dataset.}

In this work we make use of galaxy mock catalogs from the \sancho suite\footnote{Documentation and free access to the full suite is found \href{https://quijote-simulations.readthedocs.io/en/latest/sancho.html}{here}.}. The complete \sancho data-set encompasses $240$,$000$ galaxy mocks in redshift space scanning across $11$ cosmologies, $3$ massive neutrino cosmologies, $6$ primordial non-Gaussianity amplitudes, and $11$ Halo Occupation Distribution (HOD) models. These catalogs are based on the \quijote N-body simulations \citep{VillaescusaNavarroEtal2020}, run with $512^3$ dark matter particles in a box of volume $1\cGpc$. Galaxies are populated on halo catalogs using a $5$ parameter Halo Occupation Distribution (HOD) model, as described in \citep{ZhengCoilZehavi2007}. The HOD parameters are chosen such that the catalogs mock the CMASS galaxies observed by BOSS \citep{BeutlerEtal2017}, the average number density being $\bar n_{\rm{g}} \approx 3 \times 10^{-4}\,h^3\,{\rm Gpc}^{-3}$ at redshift $z=0.5$. In this work, we use a subset of almost $5$,$000$ catalogs from the fiducial cosmology set, a $\Lambda$CDM cosmology with $\Omega_{\rm{m}}=0.3175$, $\Omega_{\rm{b}}=0.049$, $h=0.6711$, $n_{\rm{s}}=0.9624$, $\sigma_8=0.834$ and $\sum m_\nu = 0.0 \,$eV. \\

\paragraph*{Measurements and covariance} The galaxy power spectrum monopole, quadrupole, and hexadecapole in redshift space ($P_\ell(k)$ with $\ell=\left\{0,2,4\right\}$) are measured using the public code \textsf{PBI4}\footnote{\href{https://github.com/matteobiagetti/pbi4}{https://github.com/matteobiagetti/pbi4}} \citep{BiagettiVerdianiSefusatti2023}. The code computes the density grid using a fourth-order mass-assignment scheme with interlacing as described in \citep{SefusattiEtal2016}. The wavenumbers bins have a width of $\Delta k = 2 k_f$, where $k_f=0.006\kMpc$ is the fundamental frequency of the \sancho box, and are computed up to $k_{\rm max} = 0.45\kMpc$. These measurements are publicly available along with the \sancho dataset. The covariance used in the analysis is estimated numerically from these measurements as
\begin{equation}
    \hat{C}_{ij}^{\ell\ell'} =  \langle [\hat{P}_\ell^{(\alpha)}(k_i) - \langle\hat{P}_\ell(k_i)\rangle]\, [\hat{P}_{\ell'}^{(\alpha)}(k_j) - \langle\hat{P}_{\ell'}(k_j)\rangle]\rangle
\end{equation}
where $\hat{P}_\ell^{(\alpha)}(k)$ is the $\alpha$-th measurement of the power spectrum multipoles, while the expectation value $\langle...\rangle$ is obtained as the average over the fiducial cosmology simulations set comprising 4,000 realisations. Given this large number and a data vector size not larger (but in most cases much smaller) than 108, we ignore statistical errors affecting the resulting covariance (see, e.g.\citep{PercivalEtal2014}). 
\section{Methodology}
\label{sec:methodology}

\subsection{Theoretical model}
\label{subsec:Theoretical model}

We consider the next-to-leading order prescription for the galaxy power spectrum multipoles in redshift space implemented in the \pbj code adopted in \citep{OddoEtal2020, OddoEtal2021, CarrilhoEtal2021, CarrilhoMorettiPortsidou2023, MorettiEtal2023a}. We refer the reader to these works for a proper description of the theoretical prediction. This corresponds to the Effective Field Theory of Large-Scale Structure model tested in the challenge paper of \citep{Nishimichi2012} and employed, for instance, in the analyses of the BOSS data in \citep{IvanovSimonovicZaldarriaga2020, DAmicoEtal2020}.
In brief, we adopt the following bias expansion up to the third order, 
\begin{equation}
\label{completup3}
    \delta_{g}=b_{1}\delta+\frac{b_{2}}{2}\delta^{2}+b_{{\mathcal{G}_{2}}}\mathcal{G}_{2}+b_{\Gamma_{3}}\Gamma_{3}+\epsilon
\end{equation}
where $b_{1}$ and $b_{2}$ are respectively the linear and quadratic bias while $\mathcal{G}_{2}$ and $\Gamma_{3}$ are two Galileon operators which quantify non local effects on the density field.
The expression for the galaxy power spectrum in redshift-space is given by 
\begin{align}
\label{eq:Ptheory}
P_{g}(\kv) &= Z_1^2(\kv) \PL(k) \nonumber \\
&+ 2 \int \de^3 \qv \, \left[ Z_2(\qv, \kv - \qv) \right]^2\PL(q) \PL(\vert \kv - \qv \vert) \nonumber \\
&+ 6 Z_1(\kv)\PL(k) \int \de^3 \qv \, Z_3(\kv, \qv, -\qv)\PL(q) \nonumber \\
&+ P_{\rm noise}(\kv)+ P_{\rm ctr}(\kv) .
\end{align}
where $Z_1(\kv)$, $Z_2(\kv_{1}, \kv_{2})$ and $Z_3(\kv_{1}, \kv_{2}, \kv_{3})$ are the usual redshift-space kernels (see, for instance, \citep{IvanovSimonovicZaldarriaga2020} for their full expression corresponding to the adopted bias expansion).
The last two terms in the eq.~(\ref{eq:Ptheory}) are the contributions from shot-noise and the EFT counterterms. The first is given by 
\begin{equation} 
\label{ShotNoisePs}
    P_{\textnormal{noise}}(k, \mu)=\frac{1}{\bar{n}}\left[(1+ \alpha_{P}+ \epsilon_{0,k^{2}}k^{2}+\epsilon_{2,k^{2}}k^{2}\mu^2)\right]
\end{equation}
where $\bar{n}$ is the (known) mean number density of objects while the parameters $\alpha_{P}$, $\epsilon_{0,k^{2}}$ and $\epsilon_{2,k^{2}}$ describe constant and scale-dependent deviations from the Poisson prediction. In the expression above and in what follows, $\mu$ represents the cosine of the angle formed by the line of sight and the wavevector $\kv$. The EFT counterterms contribution is given instead by
\begin{align}
\label{Counterterms}
 P_{\rm ctr}(k,\mu)=&-2 \tilde{c}_{0}k^{2}\PL(k)-2 \tilde{c}_{2}k^{2} f \mu^{2} \PL(k)\nonumber\\
 & -2 \tilde{c}_{4}k^{2} f^{2} \mu^{4} \PL(k) \nonumber\\
 &+c_{\nabla^{4}\delta}(\mu k f)^{4}(b_{1}+f\mu^{2})^{2}\PL(k).
\end{align}
The power spectrum multipoles are then defined as
\begin{equation}
P_{\ell}(k)\equiv \frac{(2l+1)}{2}{\int{d\mu} \ P_{g}(\kv)\mathcal \ {P}_{l}(\mu)},  
\end{equation}
where $\ {P}_{l}(\mu)$ are the l-order Legendre polynomials.

In order to accurately describe the nonlinear evolution of the acoustic features we implement the Infrared Resummation \citep{SenatoreZaldarriaga2014} following \citep{BaldaufEtal2015B} with the wiggle-no wiggle split from the 1D Gaussian filter method described in \citep{VlahEtal2016}. 

The model depends on 7 bias and shot-noise parameters plus 4 counterterms, in addition to the cosmological parameters. These are reduced to 6 and 3 when only monopole and quadrupole are considered. 

Given the large number of fits we perform, we fix all cosmological parameters to their fiducial value: this allows us to avoid re-computing the linear power spectrum and the PT contributions at each step in the Markov chain, leading to a significant speed up in the analysis. However, we keep as a free parameter the growth rate parameter $f$, to be used as a proxy for the cosmological parameters in the definition of the figure of bias. 

Since we explore only parameters that can be factorised in each contribution to the full power spectrum multipoles, we can also account exactly for the effect of wavenumber bins adopted by the estimator. We do so by considering the average of the model over the discrete wavenumbers falling into each $k$-bin, following the procedure described in section 3.2 of \citep{OddoEtal2021}. 

\subsection{Likelihood analysis}

We assume a Gaussian likelihood for the data vector corresponding to the three power spectrum multipoles 
$\left\{P_0(k_1),...,P_2(k_1),...,P_4(k_1),...\right\}$
where each observable is limited in range by a chosen $k_{\rm max}$.

For each measurement $\alpha$ we define the log-likelihood 
\begin{align}
    \ln {\mathcal L}_\alpha\equiv &-\frac12\chi_\alpha^2\nonumber \\
    \equiv & -\frac12\sum_{\ell,\ell'}\sum_{k_i,k_j}^{\kmax}\delta P_\ell^{(\alpha)}(k_i)\,[\hat{C}^{-1}]_{ij}^{\ell\ell'}\,\delta P_{\ell'}^{(\alpha)}(k_j)\,,
\end{align}
where $\delta P_\ell^{(\alpha)}(k_i)\equiv\hat{P}_\ell^{(\alpha)}(k_i)-P^{\rm th}_\ell(k_i)$ and $[\hat{C}^{-1}]_{ij}^{\ell\ell'}$ denotes a generic element of the inverse of the full covariance matrix for the data vector obtained by joining the three multipoles. 

The log-likelihood corresponding to a set of $N_R$ data vector measurements is given by the sum
\begin{equation}
    \ln {\mathcal L}_{\rm tot}=\sum_{\alpha=1}^{N_R}\,\ln {\mathcal L}_\alpha=-\frac12\sum_{\alpha=1}^{N_R}\,\chi^2_{\alpha}\,.
\end{equation}

We adopt uniform priors for all parameters, as detailed in table~\ref{tab:priors}.

\begin{table} 
\centering
\begin{tabular}{ |p{2cm}||p{2cm}| }
 \hline
Parameter &  Prior \\[2pt]
 \hline 
 $b_{1}$& $\mathcal{U}(0.9,3.5)$ \\[2pt]
$b_{2}$& $\mathcal{U}(-4,4)$ \\[2pt]
$b_{{\mathcal{G}_{2}}}$& $\mathcal{U}(-4,4)$ \\[2pt]
$b_{\Gamma_{3}}$& $\mathcal{U}(-10,10)$ \\[2pt]
$\alpha_{P}$& $\mathcal{U}(-1,1)$ \\[2pt]
$\epsilon_{k^{2}}$& $\mathcal{U}(-10,10)$ \\[2pt]
$\epsilon_{k^{4}}$& $\mathcal{U}(-100,100)$ \\[2pt]
$c_{0}$& $\mathcal{U}(-100,100)$ \\[2pt]
$c_{2}$& $\mathcal{U}(-500,500)$ \\[2pt]
$c_{4}$& $\mathcal{U}(-500,500)$ \\[2pt]
$c_{\nabla^{4}\delta}$& $\mathcal{U}(-500,500)$ \\[2pt]
$f$& $\mathcal{U}(0.1,1)$ \\[2pt]
 \hline
\end{tabular}
\caption{\label{tab:priors} Uniform priors for the model parameters adopted in the analysis.}
\end{table}

The evaluation of the posterior distribution has been performed by running Monte Carlo Markov Chain (MCMC) simulations with the affine invariant sampler implemented in the \texttt{emcee} package \citep{ForemanMackeyEtal2013}.
We choose to run the MCMC with $50$ walkers and $50000$ steps. We ensure the convergence stopping the chains after $50$ autocorrelation times.

\subsection{Diagnostics}

To determine the range of validity of the model, parametrized by the maximum wavenumber $\kmax$, we consider three different diagnostics:  the goodness of fit (GoF), a figure-of-bias (FoB) quantifying systematic errors in the recovery of the fiducial value for a cosmological parameter and an additional consistency test. We describe them below.\\ 

\paragraph*{Goodness of fit}
For the evaluation of the goodness of fit we compute the reduced chi-square statistic $\chi^{2}_{\rm red}=\chi^{2}_{\rm tot}/N_{\rm dof}$, $N_{\rm dof}$ being the number of degrees of freedom. The $\chi^2_{\rm tot}$ is computed as an average over the posterior. If, for a given value of $\kmax$, the reduced chi-square falls outside the $95\%$ confidence interval expected for the value of $N_{\rm dof}$, the model ``fails'' and we set the reach of the model equal to the previous $\kmax$ value.\\

\paragraph*{Figure of bias (FoB)}
With the figure of bias we estimate the size of the systematic error induced by the model on the cosmological parameters, as a function of $\kmax$. As mentioned already, in this work we limit our attention to the growth rate parameter $f$, since the analysis of the full set of cosmological parameters, to the extent of the one conducted here, would be computationally too demanding. We define the figure of bias for the parameter $f$ as \begin{equation}
    \label{Fob}
    \mathrm{FoB}= \frac{|\hat{f}(\kmax)-f_{\rm fid}|}{\sigma_{f}(\kmax)},
\end{equation}
where $f_{\rm fid}$ is the value of $f$ corresponding to the fiducial cosmology while $\sigma_f$ is the marginalised error on $f$ at that $\kmax$ value. The reach of the model based on the FoB, $k^{\rm FoB}_{\rm reach}$, is then defined by the largest value of  $\kmax$ such that $\mathrm{FoB}<2$.\\

\paragraph*{Consistency}
Finally we consider a ``consistency'' criterion based on the fact that all model parameters are constant and therefore not expected to depend on the scale $k_{\rm max}$, since this would indicate a clear failure of the model. We identify a possible running of a parameter by performing a fit of all its values at different $k_{\rm max}$ with a constant function. We then compute the reduced chi-square of the fit and compare it with the $95\%$ confidence interval, determining for each parameter the reach of the model in the same way as for the goodness of fit. The reach of the model based on the consistency criterion will then be the minimum $k_{\rm reach}$ across the different parameters.\\

It is clear that the definition of the range of acceptable values for each of these statistics is somewhat arbitrary. We made the choice of having our three definitions correspond to a 95\% probability interval, in order to have, to some extent, more homogeneous results. A looser definition (i.e. a larger confidence interval) on the model reach, for all diagnostics, lead to qualitatively similar results with higher values for the median $k_{\rm reach}$, as expected, but comparable scatter.

\section{Results and conclusions}\label{sec:results}

\subsection{Model reach and its uncertainty}

\begin{table}
\centering
 \begin{tabular}{l|lllllllll}
  \hline
  $N_{\rm sub}$ & 
    24 &
  48 & 
  72 & 
  88 & 
  200 & 
  306 & 
  400 & 
  580 & 
  1160 \\
$V_{\rm tot}$  & 166 & 83 & 55 & 45 & 20 & 13 & 10 & 6 & 3\\
  \hline
 \end{tabular}
 \caption{\label{tab:subsets} Simulations subsets. Here $N_{\rm sub}$ is the number of realisations subsets for a given cumulative simulation volume $V_{\rm tot}$ (in $\cGpc$). The value $V_{\rm tot}/(\cGpc)$ corresponds as well to the number of simulations in each subset, $N_{\rm sim}$, since each individual simulation has a volume of $1~\cGpc$. }
\end{table}

We investigate the reach of the model as a function of the size of statistical errors affecting the power spectrum measurements. We do so by considering increasingly large subsets of distinct measurements corresponding to twelve values for the cumulative simulation volume from 3 to 160$\cGpc$, as detailed in table \ref{tab:subsets}. From each set characterized by a given cumulative volume we extract three values of $k_{\rm reach}$ according to the different definitions, while we obtain an estimate of its error from the scatter across all subsets of the same size. We consider $k_{\rm max}$ values in the range $\kmax\in \left\{0.04,0.44\right\} \kMpc$ separated by the interval $\Delta k_{\rm max} = 0.04 \kMpc$ \footnote{We made some attempt to consider as well larger volumes, for instance comparable to the one characterising the challenge paper of \citep{NishimichiEtal2020}, but the limited number of subsets lead to very noisy results. In addition, the lower reach would require more $\kmax$ values to have a proper assessment of the dependence. More importantly, comparison to lower volume results becomes non-trivial, given that some parameters become unconstrained for low reaches. }.

\begin{figure}[t!]
\includegraphics[width=0.98\columnwidth]{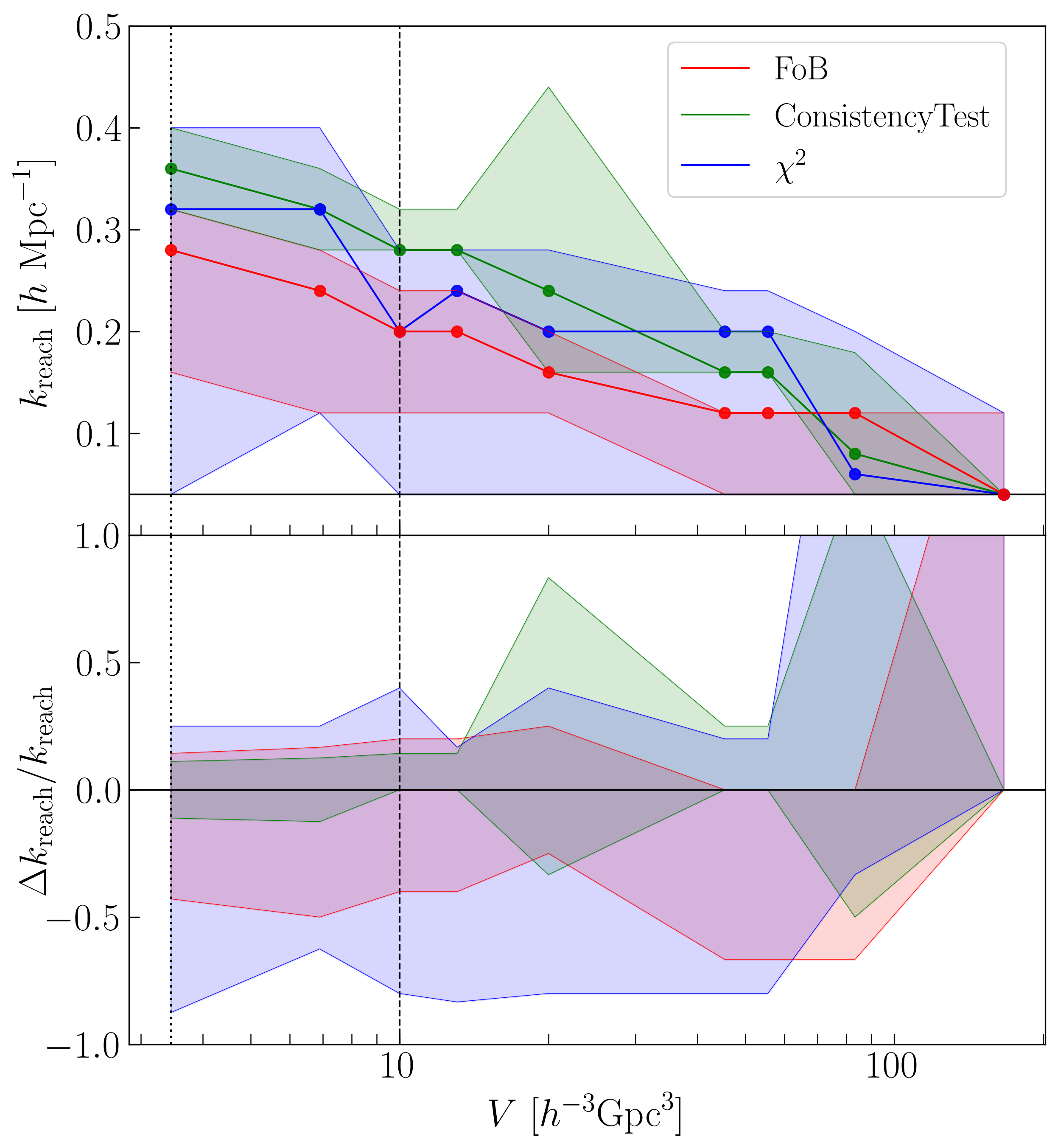}
    \caption{In the top panel we show the median of the $k_{\rm reach}$ with the inclusion of the hexadecapole in the analysis, estimated with the three different methods, as a function of the cumulative volume $V$.
    The shaded areas represent the 16th percentile and the 84th percentile.
    In the bottom panel we show the relative scatter with respect to the median, for the three methods. The two vertical lines represent the volume of a typical redshift bin for BOSS (dotted line) and a Euclid-like survey (dashed line).} 
    \label{fig:ManyplotSancho000}
\end{figure}

In the top panel of figure \ref{fig:ManyplotSancho000} we show the median of the reach of the model for the joint analysis of monopole, quadrupole and hexadecapole as a function of the cumulative simulation volume $V$. The blue, red and green points denote the $k_{\rm reach}$ defined, respectively, from the goodness of fit, the figure of bias and from the model consistency test. The shaded area fills the interval between the 16th and 84th percentile in the distribution of values across the distinct subsets of equal volume. The bottom panel shows the relative scatter with respect to the median.  The two vertical gray lines from the left mark, approximately, the volume of a typical redshift bin of the BOSS CMASS sample (3$\cGpc$, \citep{BeutlerEtal2017B}) and of the Euclid spectroscopic sample (10$\cGpc$ \citep{EuclidIST:F2019}).

As expected, the $k_{\rm reach}$ decreases as the volume increases, that is with decreasing statistical errors. However, more interesting is the scatter measured on each $k_{\rm reach}$ estimates. As evident from the lower panel, the relative scatter is quite large, up to 50\% or even 70\%, depending on the $\kreach$ definition, and approximately constant for realistic volumes below 100$\cGpc$. 

In addition we  notice how the $\kreach$ defined in terms of the goodness-of-fit criterion ($\chi^2$) has a larger scatter, and, because of this, it is largely compatible with the other two definitions. On the other hand, the one derived in terms of the figure-of-bias is consistently lower  than the $\kreach$ from the consistency test even accounting for the relative scatter. This could have relevant consequences since the second one is the criterion on which a scale-cut is determined in the analysis of actual data: if $\kmax$ is indeed chosen in this way, we can expect a significant bias in the determination of the growth rate.

\begin{figure}[t!]
\includegraphics[width=0.98\columnwidth]{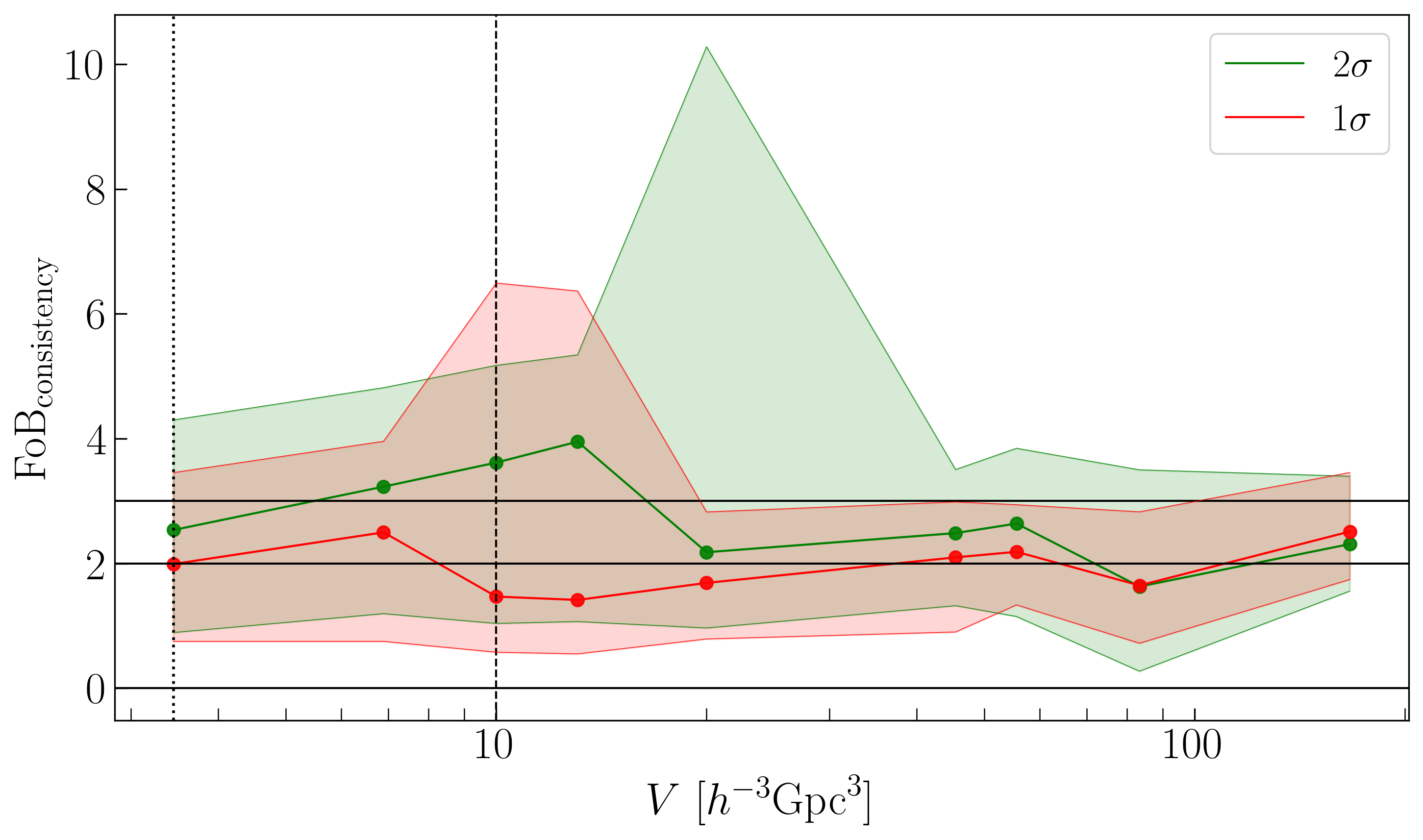}
    \caption{Median of the Fob considering the posterior of the MCMC runs with the choice $k_{\rm max}$ of the consistency method, evaluated respect to the $95 \%$ (green) and $68 \%$ (red) confidence level, as a function of the cumulative volume $V$.
    The shaded areas represent the 16th percentile and the 84th percentile.
 The two vertical lines represent the volume of a typical redshift bin for BOSS (dotted line) and a Euclid-like survey (dashed line).} 
    \label{fig:FobConsistency}
\end{figure}

We explore this aspect further in figure \ref{fig:FobConsistency}, where we plot the median and the relative scatter of the FoB evaluated over the posterior distribution for the parameter $f$ considering the $\kmax$ determined from the consistency criterion, as one could do in the analysis of real data.
In particular the red and green line represent respectively the choice of $\kmax$ with respect to the $95 \%$ and $68 \%$ confidence level of the consistency criterion. This test shows how the selection of a scale cut based on the consistency of the model parameters might lead to a significantly biased estimation of the growth rate parameter. For a consistency test that allows for a 95\% CL interval for a constant fit to each model parameter, the bias on $f$ can have a median around $3 \sigma$, i.e. most of the runs present a bias larger than 3$\sigma$. Restricting the consistency test to  a more conservative $68\%$ CL interval, the median FoB value is now close to 2$\sigma$. Yet, also in this case the scatter associated to the FoB is still rather wide with respect to its median values.

It is also interesting to notice that the theory consistency test is failed first by the $c_{\nabla^4\delta}$ counterterm, describing higher-order Finger-of-God effects. This term is strictly speaking a two-loop correction and, for this reason, is not considered by the West-Coast team in the challenge of \citep{NishimichiEtal2020}. The strong running affecting this parameter in our test could be a hint that, indeed, a consistent modelling requires the missing 2-loop corrections. We are not exploring here any of such choices, nor other descriptions (e.g. \citep{SanchezEtal2017b}) of redshift-space distortions, leaving this for future work.

\subsection{Hexadecapole information}

The overall picture described in the previous section does not change significantly when the
hexadecapole is excluded from the analysis. The median $k_{\rm reach}$ is not noticeably different, and not consistently larger either, indicating the relatively small constraining power this observable has on the model. 

The results above, however, allow to investigate more in detail the additional information provided by the hexadecapole, at least on the growth rate $f$, and how this is affected by statistical errors. To this end, we define a Figure of Merit (FoM) on such parameter given simply by the inverse of its marginalised uncertainty 
\begin{equation}
    {\rm FoM}=\frac{1}{\sigma_{f}(\kmax)}\,.
\end{equation}
We consider MCMC runs with the conservative choice $\kmax = k_{\rm reach}^{\rm FoB}$ and extract the uncertainty on the parameter $f$ from its posterior distribution. As a final result, we can obtain the figure of merit as a function of the cumulative volume.

\begin{figure}
	\includegraphics[width=0.98\columnwidth]{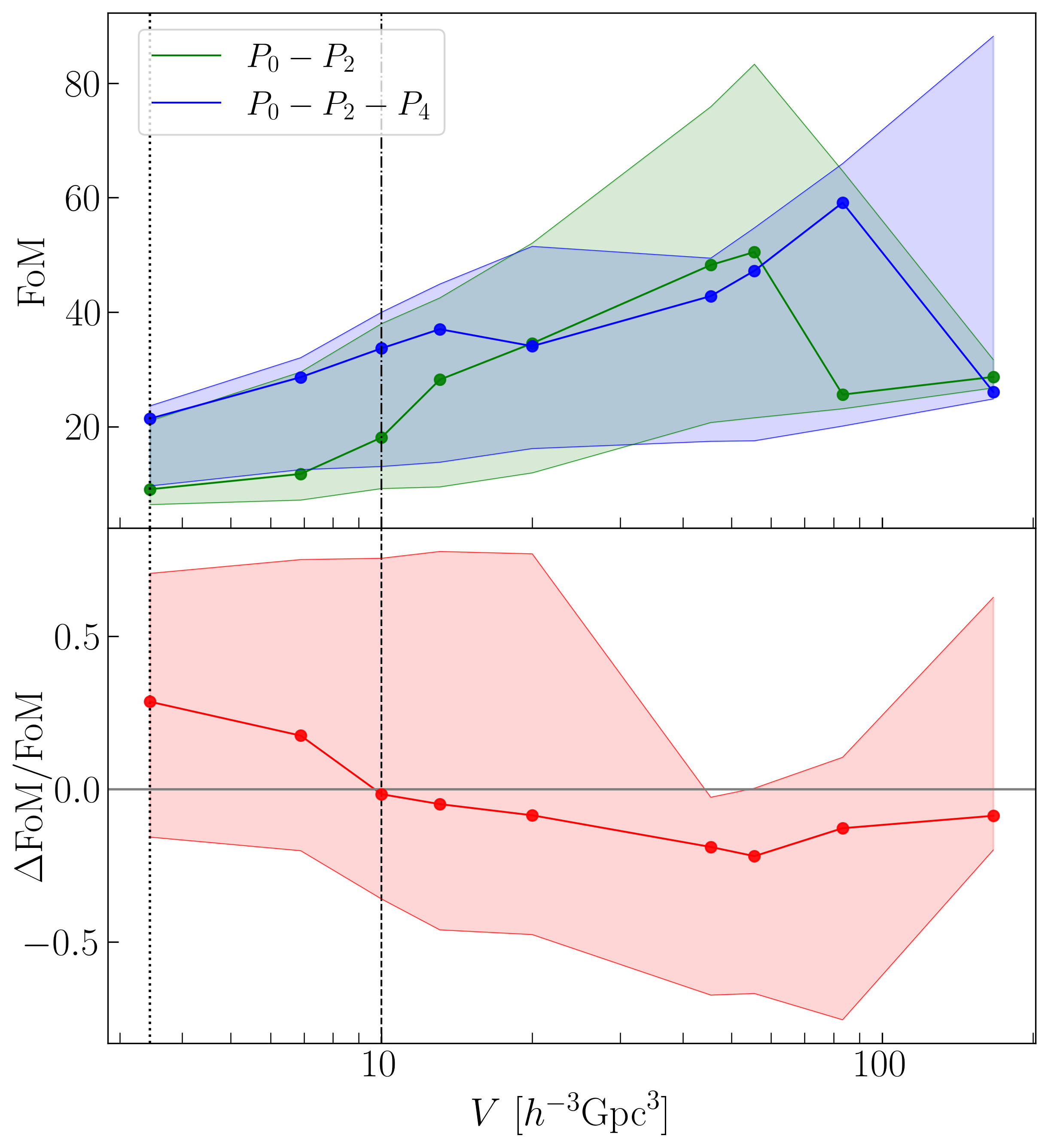}
    \caption{The top panel shows the median of the Figure of Merit, with the inclusion of the hexadecapole (bue) and without it (green),  as a function of the  volume $V$.
    In the bottom panel we plot the relative median of the differences of the Figures of Merit, with the inclusion of the hexadecapole and without it, against the volume $V$. 
    The shaded areas represent the 16th percentile and the 84th percentile.
    The two vertical lines represent respectively the volume of a typical redshift bin for BOSS (dotted) and a Euclid-like survey (dashed).}
    \label{fig:Fom}
\end{figure}

The top panel of figure \ref{fig:Fom} shows the median value of the figure of merit across all simulation subsets of equal volume along with its scatter for both analyses, including and excluding the hexadecapole information. We see how, as we can expect, the FoM grows as the volume increases, although such increase tends to flatten at the highest volumes. This is likely due to the reduced range of the analysis for these configurations. The scatter also grows significantly with the volume. Remarkably, the difference induced by the addition of the hexadecapole is quite small, and significantly smaller than the scatter. This also makes not too relevant the relatively larger FoM for the $P_0-P_2$-only analysis for some cases at large cumulative volume. 

This is more clearly shown in the bottom panel of figure \ref{fig:Fom}, where we plot the median relative difference due to $P_4$ evaluated for each subset. The improvement is largely consistent with zero for all volumes considered.

These results suggest that the hexadecapole does not provide, {\em on average}, additional information on the growth rate parameter. It is still true that, in a small but sizeable fraction of realisations close to the characteristic volume of current redshift surveys, the improvement on the FoM could be as high as $\sim$50\%.

\subsection{Discussion}

The determination of conservative scale-cuts in the analysis of observational data-sets is quite an important decision to be made, as it is necessary to avoid introducing systematic errors due to the limitations of the theoretical model. On the other hand, it is also crucial to take advantage of the available information as much as possible. 

In principle a scale-cut can be defined from a proper estimate of the theoretical error (see, e.g. \citep{BaldaufEtal2016}) compared to a numerical or analytical estimate of the statistical uncertainty. Unfortunately, theory errors could not be determined so far with sufficient accuracy to be used in practice for this purpose.    

We provided here a first simple, but extensive, assessment of the statistical errors affecting the determination of the validity range of the theory according to three criteria. In the analysis of actual data, of course, the FoB considered here cannot be used for the determination of the scale cuts, while the consistency test is the typical, practical choice. We find, however, that the $k_{\rm reach}$ determined with the latter is consistently higher than the one from the FoB, and it could therefore allow some level of systematic error on the determination of cosmological parameters. Moreover, it is also characterised by a scatter with a significant variability across the cumulative volumes considered here. 
We are well aware that our work is limited in many ways. It focuses on a single parameter, $f$, as a proxy for the whole set of cosmological parameters. It considers a single theoretical model with a given choice of priors and does not explore possible projections effects (see, e.g. \cite{CarrilhoMorettiPortsidou2023}) while it would be interesting to see how a comparison between different theoretical assumptions or phenomenological models would look like when conducted on a large subsets of catalogs with variable cumulative volume.

Even our assessment of the amount of information in the power spectrum hexadecapole could benefit from a more general likelihood analysis, perhaps considering alternative statistics to limit FoG effects \citep{2021arXiv211000016D, IvanovEtal2022}.

Yet, while we leave these possible explorations to future work, we believe this simple exploration provides an interesting first estimates of the uncertainty affecting the reach of perturbative modelling. 

\section*{Acknowledgements}

We are grateful to Pierluigi Monaco, Cristiano Porciani and Marko Simonovi\'c for helpful discussions and valuable suggestions. G.G. acknowledge support from the Spanish Ministerio de Ciencia e Innovacion (MICINN), project PRE2022-102836. M.B. is supported by the Programma Nazionale della Ricerca (PNR) grant J95F21002830001 with title "FAIR-by-design". C.M. is supported by the Italian Research Center on High Performance Computing Big Data and Quantum Computing (ICSC), project funded by European Union - NextGenerationEU - and National Recovery and Resilience Plan (NRRP) - Mission 4 Component 2 within the activities of Spoke 3 (Astrophysics and Cosmos Observations).

\bibliographystyle{apsrev}
\bibliography{cosmologia}

\end{document}